\RequirePackage{fix-cm}

\documentclass[%
floatfix,
showkeys,
twocolumn, %
nofootinbib, %
superscriptaddress, %
]{revtex4-1}

  \usepackage{cmap}
\usepackage{textcomp}

\usepackage[T1,T2A]{fontenc}

\usepackage[utf8]{inputenx}
\input{ix-utf8enc.dfu}

\usepackage[german,russian,english]{babel}

\usepackage{amsmath}
\usepackage{amssymb}

\usepackage{mathtools}
\mathtoolsset{
showonlyrefs,
mathic = true
}

\allowdisplaybreaks

\usepackage{hyperref}
\hypersetup{backref,
 colorlinks=false}
\hypersetup{pdfborder=0 0 0}

\usepackage{microtype}
\UseMicrotypeSet[protrusion]{alltext}

\usepackage{graphicx}

\usepackage[scanall]{psfrag}

\usepackage{listings}
\usepackage{listingsutf8}
\usepackage[frozencache]{minted}

\usepackage{fixltx2e}

\usepackage{nicefrac}

\makeatletter
\def\ps@pprintTitle{%
     \let\@oddhead\@empty
     \let\@evenhead\@empty
     \let\@oddfoot\@empty
     \let\@evenfoot\@oddfoot}
\makeatother

\begin{document}
\graphicspath{{image/rng-cas-sympy/en/}{image/rng-cas-sympy/}{image/}}

  \title{A modular extension for a computer algebra system}

\author{Migran N. Gevorkyan}
\email{gevorkyan-mn@rudn.ru}
\affiliation{Department of Applied Probability and Informatics,\\
  Peoples' Friendship University of Russia (RUDN University),\\
  6 Miklukho-Maklaya St, Moscow, 117198, Russian Federation}

\author{Anna V. Korolkova}
\email{korolkova-av@rudn.ru}
\affiliation{Department of Applied Probability and Informatics,\\
  Peoples' Friendship University of Russia (RUDN University),\\
  6 Miklukho-Maklaya St, Moscow, 117198, Russian Federation}

\author{Dmitry S. Kulyabov}
\email{kulyabov-ds@rudn.ru}
\affiliation{Department of Applied Probability and Informatics,\\
  Peoples' Friendship University of Russia (RUDN University),\\
  6 Miklukho-Maklaya St, Moscow, 117198, Russian Federation}
\affiliation{Laboratory of Information Technologies\\
  Joint Institute for Nuclear Research\\
  6 Joliot-Curie, Dubna, Moscow region, 141980, Russia}

\author{Leonid A. Sevastianov}
\email{sevastianov-la@rudn.ru}
\affiliation{Department of Applied Probability and Informatics,\\
  Peoples' Friendship University of Russia (RUDN University),\\
  6 Miklukho-Maklaya St, Moscow, 117198, Russian Federation}
\affiliation{Bogoliubov Laboratory of Theoretical Physics\\
  Joint Institute for Nuclear Research\\
  6 Joliot-Curie, Dubna, Moscow region, 141980, Russia}

\begin{abstract}
Computer algebra systems are complex software systems that cover a wide range of scientific and
practical problems. However, the absolute coverage cannot be achieved. Often, it is required to create a user
extension for an existing computer algebra system. In this case, the extensibility of the system should be taken
into account. In this paper, we consider a technology for extending the SymPy computer algebra system with
a low-level module that implements a random number generator.
\end{abstract}

\maketitle

\section{Introduction}

The SymPy system~\cite{lamy:sympy_starter} is essentially a module
written in Python~\cite{slatkin:book:effective-python::en,lubanovic:book:introducing-python::en}.
Hence, to extend the functionality of SymPy, it is sufficient to
write a function or module in Python.

Even though Python is a universal programming language that can be
used to implement any algorithm, it has a significant drawback: low
performance. Performance drop is especially noticeable when the
algorithm contains loops. This problem is due to the dynamic nature of
the language: even elementary data types, e.g., int and float, are
implemented in the standard interpreter cpython as composite data
structures.

Python is well suited for prototyping applications. In addition,
Python is a kind of glue: it can be used for linking different
libraries~\cite{kulyabov:2019:programmirovanie:cadabra::en}. However,
an attempt to write a large, fast program in this language is likely
to fail. The reason why Python is so successful in solving scientific
and engineering problems is that it uses a low-level interface to the
libraries written in more computationally efficient programming
languages.  This gives the impression that Python is as fast as the
code written, e.g., in C++. As a result, for practical programming in
Python, one should also learn lower-level languages. If standard
libraries are sufficient to solve a problem, then Python may be the
only tool.  However, if we need to add new functionality, then we have
to employ lower-level programming languages.  This paper discusses the
use of the Python module ctypes for the integration of C functions
into a Python program. As an example, we consider a library that
implements a random number generator. This example is interesting
because it is quite resource-intensive and impractical to implement in
pure Python.

It should be noted that the idea of using a compiled statically-typed
language to improve the performance of individual components of a
program written in an interpreted dynamically-typed language is not
new~\cite{aladjev:book:maple,corless:book:essential-maple-7}. On the contrary, there are many technological approaches
that enable this extension. Moreover, the variety of these approaches
makes it difficult for novice programmers to enter this area. In this
paper, we present a brief overview of approaches to improving the
performance of Python programs.

The main focus is placed on using the built-in module ctypes for the
direct call of C functions from Python programs. The choice of this
module is substantiated. The discussion is practice focused and
addresses the development of a library in C, as well as its
compilation for further use with ctypes on various platforms. That
part of the paper can be used as an introduction to ctypes for
beginners. The last part of the paper describes our module, which uses
ctypes to call functions from the C library that implements a number
of pseudorandom generators. We also present some test results to
compare the proposed pseudorandom generator module with the generators
from the random module and NumPy library.

\section{Improving the performance of sympy/python programs}

Below we list the basic approaches generally
employed to boost Python programs.

\begin{itemize}

\item The use of third-party libraries, e.g.,
  NumPy~\cite{idris:numpy_cookbook,oliphant:guide_numpy} and
  SciPy~\cite{oliphant:2007:scipy}, in which resource-intensive
  algorithms are implemented in C/C++ and Fortran, while Python is
  used as a glue language to provide a convenient programming
  interface.

\item The optimizing static compiler
  Cython~\cite{behnel:2011:cython,smith:book:cython}, which allows
  Python code to be translated into C/C++ code. In this case, the
  program itself is written in a special dialect of Python, which
  enables static typing in performance-critical sections of the code.

\item The Numba project~\cite{lam:2015:numba-llvm}, which is a JIT compiler for
  Python. Numba allows one to write programs in pure Python by using
  decorators for functions and loops the performance of which needs to
  be improved.

\item The module
  ctypes~\cite{spreitzenbarth:book:mastering-python-forensics} from
  the standard Cpython interpreter library, which allows C functions
  to be directly called from static or shared libraries just as
  ordinary Python functions.
\end{itemize}

None of these approaches is universal. The specialized libraries NumPy
and SciPy are focused mostly on
scientific and engineering computational problems,
which is why their sets of algorithms, although being
extensive, are restricted by their specialization. Numba
(JIT compiler) provides a significant increase in
speed; however, this project is still under development
and its functionality is limited by standard application
scenarios.

Cython (static compiler) is by far one of the most popular tools for
program speedup. It is employed in many libraries, including NumPy and
SciPy, to improve the performance and integration of libraries in
C/C++.

In this paper, we use the ctypes module as it has several advantages
over Cython that are important for our research:
\begin{itemize}
\item it is a standard Cpython module, while Cython must be installed
  separately;

\item it does not mix several dialects of Python;

\item it is especially useful when the required functionality is
  already implemented in C; in this case, the implementation of a
  function call is extremely easy and takes just a few lines of
  code. It is also useful when implemented C functions are simple but
  actively use loops and primitive data types.
\end{itemize}

Thus, with ctypes, programming is divided into two steps. At the first
step, the programmer implements functions in C and compiles them to
assemble a static or shared (dynamic) library. The second step
involves separate implementation of some wrapper functions in
Python. These wrapper functions provide an interface for the
convenient call of the implemented C functions from Python
programs. Note that the C function call functionality is implemented
using the ctypes module included in the standard CPython interpreter
library.

\section{Ctypes module}

This section describes the full cycle of developing a
C library and its use in combination with ctypes. The
official documentation on ctypes contains some examples of using
functions from this module; however, it
does not cover the development of libraries in C.

The use of the ctypes module begins with loading a
library file. That is why, before proceeding to the basic
functionality of ctypes, we consider an example of
assembling static and dynamic (shared) libraries by
using the C compiler from the gcc compiler set. For
this purpose, we used gcc 8.3.0 under GNU Linux
(Ubuntu 19.04).

\subsection{Compiling C functions and assembling the library}

The code of a typical C library consists of a set of
source code files (\mintinline{bash}|crc_01.c|, \mintinline{bash}|crc_02.c|, etc.) and
a number of header files (\mintinline{bash}|header_01.h|, etc.).
Common practice~\cite{klemens:book:21st-century-c::en} is to place all source files in the
subdirectory \mintinline{bash}|src| of the project and header files in the
\mintinline{bash}|include| subdirectory.

To compile the source files, the following compiler
keys are used.
\begin{itemize}
\item \mintinline{bash}|-c|: allows one to create an object file without
assembling the whole program or library.

\item \mintinline{bash}|-Wall|: in addition to messages about syntax
errors, the compiler prints warnings that can poten-
tially lead to an incorrect operation of the program.

\item \mintinline{bash}|-Werror|: the compiler interprets all warnings
as errors.

\item \mintinline{bash}|-fPIC|: tells the compiler to translate the pro-
gram into position-independent code (PIC), where all
jumps are made only at relative addresses. This flag is
important because the library can be loaded anywhere
in the program.

\item \mintinline{bash}|-I./include|: tells the compiler to search for
header files in the local include directory of our
project.

\item \mintinline{bash}|-L./lib|: tells the compiler to search for library
files in the local lib directory of our project. It is
important to follow the sequence and set the \mintinline{bash}|-L| flag
only after the \mintinline{bash}|-I| flag.
\end{itemize}

Upon debugging the program, one can also add
the optimization flag: \mintinline{bash}|-O2| or \mintinline{bash}|-O3|. It should be noted,
however, that aggressive optimization in some cases
can cause incorrect operation of functions.

All flags listed above are stored in the environment
variable \mintinline{bash}|CFLAGS|. To compile a source file into an
object file, the following command is executed for
each file:  
\begin{minted}[autogobble,breaklines=true,breakanywhere]{bash}
    gcc -c $CFLAGS src/src_01.c -o lib/obj_01.o
\end{minted}

Once all object files are created, they can be packed
into a static library by using the ar utility; for this purpose, the
following command is executed:
\begin{minted}[autogobble,breaklines=true,breakanywhere]{bash}
    ar crs libmy.a lib/obj_01.o lib/obj_02.o
\end{minted}

The \mintinline{bash}|crs| option means that it is required to create
an archive with the replacement of the files, if it contains any, and
also create an index. The successful execution of the command yields a
static library file
\mintinline{bash}|libmy.a|, which can be used for connection by
means of \mintinline{bash}|ctypes|.

To create a static library in Windows, the \mintinline{bash}|-Wl|
option is used, which allows one to pass additional
options to the linker and specify (using the linker
option \mintinline{bash}|--out-implib|) the path to the file of the
static library that needs to be created.
\begin{minted}[autogobble,breaklines=true,breakanywhere]{bash}
    gcc -shared lib/obj_01.o lib/obj_02.o libmy.so -o bin/libmy.dll -Wl,--out-implib,lib/win/libmy.a
\end{minted}

If necessary, a shared library (rather than a static
one) can be created:
\begin{minted}[autogobble,breaklines=true,breakanywhere]{bash}
    gcc -shared lib/obj_01.o lib/obj_02.o libmy.so
\end{minted}
This command yields a shared library file
\mintinline{bash}|libmy.so|. The same command allows one to create
a dynamic library in Windows. It is only required to
specify the file extension \mintinline{bash}|.dll| instead of \mintinline{bash}|.so|.

\subsection{Loading the library into the python program}

Assuming that we have successfully compiled and
assembled the library \mintinline{bash}|libmy.so|, we can proceed to
its import into the Python program. Suppose that the
library file is in the same directory as our Python program. Let us
consider the following code fragment.

\begin{minted}[autogobble,breaklines=true,breakanywhere]{python}
    import ctypes
    import sys
    import os
    
    path = os.path.dirname(os.path.realpath( __file__ ))

    if sys.platform.startswith('win'):
        clib = ctypes.CDLL(os.path.join(path, 'libmy.dll'))
    else:
        clib = ctypes.CDLL(os.path.join(path, 'libmy.so'))
\end{minted}

First, the \mintinline{python}|ctypes| module and some additional
modules need to be loaded. Next, we obtain the absolute path to the
directory of the program. Then, we
determine the type of the operating system and,
depending on it, load the \mintinline{bash}|.dll| or \mintinline{bash}|.so| file.

It should be noted that the library file must be
loaded by the absolute path if we organize our Python
program as a module and want to store the library file
in the module's directory.

\subsection{Calling functions form the library}

Once imported, all library functions are available
for calling as attributes of the clib object. Suppose
that the libmy library contains the following function:

\begin{minted}[autogobble,breaklines=true,breakanywhere]{c}
uint64_t uint64_var(uint64_t var) {
    uint64_t i = 9223372036854775807llu;
    printf("Function uint64_var, arg uint64 var = %llu\n", i);
    return i;
}
\end{minted}
To call it from the Python program, the following
code is used:
\begin{minted}[autogobble,breaklines=true,breakanywhere]{python}
    clib.uint64_var.argtypes = [ctypes.c_uint64]
    clib.uint64_var.restype = ctypes.c_uint64
    res = clib.uint64_var(0)
\end{minted}
Before calling the function, the type of its argument is specified
using a single-element list because it
has only one argument. Next, the type of the return
value is specified; once this is done, the function can
be called. Ctypes defines all standard types of C, and
these three lines of code cover all instances of calling
any simple function that receives and returns arguments of basic types.

Let us consider a more complex example where the
argument is passed to the function by a pointer. Sup-
pose that we have the following C function:
\begin{minted}[autogobble,breaklines=true,breakanywhere]{c}
void change_var(double* var) {
    *var = 2.0;
}
\end{minted}

The following code illustrates a way to call this
function by using \mintinline{python}|ctypes|:
\begin{minted}[autogobble,breaklines=true,breakanywhere]{python}
    x = ctypes.c_double(1.0)
    print(f"x = {x.value}")
    clib.change_var(ctypes.byref(x))
    print(f"x = {x.value}")
\end{minted}

Here, we first assign the value of the variable \mintinline{c}|x| by
using the constructor \mintinline{c}|c_double| and then pass it as
an argument to the function \mintinline{c}|change_var| while
additionally specifying (by using \mintinline{c}|byref|) that the
argument is passed by a reference. Since the function
does not return any values, it is not necessary to specify \mintinline{c}|restype|,
and, having passed the variable of the
known type as an argument, we do not have to specify
the type of the argument.

Finally, let us consider the call of a function that
takes an array as an argument:  
\begin{minted}[autogobble,breaklines=true,breakanywhere]{c}
double avg_value(long long int array[], size_t len) {
    double avg = 0.0;
    for (size_t i = 0; i < len; ++i) {
        avg += (double) array[i] / (double) len;
        array[i] = 0.0;
    }
    return avg;
}
\end{minted}

The function \mintinline{c}|avg_value| takes an array as the
first argument and an integer (the size of the array) as
the second argument. To facilitate the call of this function from the
Python code, we can write the following
wrapper function.  
\begin{minted}[autogobble,breaklines=true,breakanywhere]{python}
def avg_value(l: list) -> float:
    """avg_value wrapper"""
    clib.avg_value.restype = ctypes.c_longdouble
    A = (ctypes.c_longlong * len(l))(*l)
    n = ctypes.c_size_t(len(l))
    return clib.avg_value(A, n)
\end{minted}

First, the return type (\mintinline{c}|long double|) is specified;
next, memory is allocated for the array, which is
immediately initialized with values from the list \mintinline{c}|l|.
Then, the variable \mintinline{c}|n| of the type \mintinline{c}|size_t| is created,
and the C function is called.

Using wrapper functions is justified in most cases as it allows one to
hide the routine operations on argument initialization and data type
specification, thus providing a user-friendly interface.

\section{Generation of pseudorandom numbers}

The generation of truly random numbers is quite a
difficult problem. For this purpose, various physical
processes are generally used. The main problem of
true random number generators is their low rate of
generation~\cite{galton:1890:dice-statistical}. Thus, in practice, pseudorandom
generators are employed~\cite{knuth:aop:2::en,
  kulyabov:2018:ceur-ws:2267:random-number}.

The SymPy package does not implement individual pseudorandom
generators because all necessary functionality is provided by the
standard module random and submodule numpy.random of the NumPy
library.

The functions of both the modules are based on the Mersenne Twister
(MT) algorithm~\cite{L_Matsumoto:1998:MTE}, which generates uniformly distributed
pseudorandom sequences of unsigned integers. This algorithm yields
high-quality sequences of pseudorandom numbers; however, it has
relatively low performance due to its awkwardness.  Presently, a
number of alternative algorithms appeared~\cite{L_xorshift:2003,L_xorshift:2005,L_Boldi2018,L_oneill:pcg2014}, which also yield
high-quality sequences of pseudorandom numbers while outperforming the
MT in speed.

Modern algorithms for pseudorandom number generation use bitwise
logical operations and shift operations, which is why system
programming languages that provide minimum abstractions in favor of
maximum performance are a natural choice for implementing these
algorithms. Most developers also implement them in C or C++.

These implementations are compact functions
with the following signature:
\begin{minted}[autogobble,breaklines=true,breakanywhere]{c}
    uint32_t generator(uint32_t seed[]);
    /* or */
    uint64_t generator(uint64_t seed[]);
\end{minted}
where the array \mintinline{c}|seed| contains the initial values for the
initialization of the generator. To generate a sequence
of pseudorandom numbers, it is sufficient to call this
function in a loop $N$ times.
\begin{minted}[autogobble,breaklines=true,breakanywhere]{c}
    void rand_n(uint64_t N, uint64_t seed[], uint64_t res[]) {
        for (uint64_t i=0; i < N; ++i) {
            res[i] = generator(seed);
        }
    }
\end{minted}
The internal state of the generator is determined by a
set of numbers \mintinline{c}|seed| and is preserved from call to call
because the \mintinline{c}|seed| array is passed by a reference.

To obtain numbers from the half-interval $[0 , 1)$, it is
sufficient to normalize the numbers generated:
\begin{minted}[autogobble,breaklines=true,breakanywhere]{c}
double normed_gen(uint64_t seed[]) {
    return (double) generator(seed) / (double) UINT64_MAX;
}
\end{minted}

\subsection{Library structure}

We implemented a compact library in C~\cite{kulyabov:2019:rng-cas-sympy-code} that
includes some modern pseudorandom generators~\cite{rose:2018:kiss,L_xorshift:2003,L_xorshift:2005,L_Boldi2018,L_oneill:pcg2014}.
The library has the following structure.
\begin{itemize}

\item The directory \mintinline{bash}|src| contains the source files of various
  pseudorandom generator algorithms (the name of the file corresponds
  to the name of the algorithm).

\item The directory \mintinline{bash}|include| contains a single header file in which all
  functions implemented in the library are declared.

\item The directory \mintinline{bash}|tools| stores a C program that implements the
  command-line utility \mintinline{bash}|random|, which is used to run any generator and
  print the resulting sequence.

\item To assemble the library and utility under Unix-like operating
  systems, a makefile was written; for Windows, a \mintinline{bash}|.bat| file is used.

\item After compilation and assembly, we obtain files of the shared
  and static libraries that are located in the directories \mintinline{bash}|lib/shared|
  and \mintinline{bash}|lib/static|, respectively. In addition, the command utility
  random is assembled in the directory \mintinline{bash}|bin|.
\end{itemize}  

\subsection{Wrapping the library by using ctypes}

The library described above is integrated into the
Python/SymPy environment by using the standard
module \mintinline{python}|ctypes| and is designed as a Python module
called \mintinline{python}|crandom|. The functionality of the module is
implemented as a class \mintinline{python}|Random| stored in the file
\mintinline{bash}|crandom.py|.

For correct operation, the standard modules \mintinline{python}|random|, \mintinline{python}|typing|, \mintinline{python}|ctypes|,
\mintinline{python}|sys|, and \mintinline{python}|os| are required.  In addition, to generate pseudorandom
arrays, the NumPy library is employed.

Let us consider the basic capabilities of \mintinline{python}|crandom|
on some examples. To run the examples, we used
Python 3.6.8 Miniconda and Jupyter 4.4.0.

The work with the module begins with the selection and initialization
of the generator. Consider the following example.
\begin{minted}[autogobble,breaklines=true,breakanywhere]{python}
import crandom
gen = crandom.Random('xorshift+')
gen.set_seed([233, 43])
\end{minted}

Here, the object \mintinline{python}|gen| is created for the algorithm
\emph{xorshift+}. In this example, the generator is also initialized by
passing it two integers with the use of the
method function \mintinline{python}|set_seed|. If the generator is not
initialized by explicitly calling \mintinline{python}|set_seed|, then the
function \mintinline{python}|randint| from the standard module
\mintinline{python}|random| is employed. In
addition, when selecting initial
values, we should adhere to certain
recommendations~\cite{L_GoodPractice:2010}, and numbers 233 and 43
were selected so as not
to overload the example.

The state of the generator is stored in the \mintinline{python}|seed|
attribute of the \mintinline{python}|gen| object. Depending on the type of
the generator, seed can be either a single unsigned
integer or a sequence of unsigned integers. When
called without arguments, the method function
\mintinline{python}|set_seed| determines the number of integers
required for initialization.

Once the \mintinline{python}|gen| object is created and initialized, it
can be used to derive a pseudorandom sequence of a
specified size. This can be done as follows:
\begin{minted}[autogobble,
  % linenos,
  breaklines=true,breakanywhere]{python}
r = gen.generate(size=10)
r = gen.generate(size=10, type=float)
r = gen(size=10)
\end{minted}

The first call yields a sequence of ten unsigned integers. During the
second call, the optional argument
\mintinline{python}|type| receives \mintinline{python}|float|, which
results in a sequence of
floating-point numbers generated on the half-interval
$[0, 1)$. Finally, the third line demonstrates that it is not
necessary to use the function method \mintinline{python}|generate|
because the class defines the method \mintinline{python}|__call__| and
the object itself can be called as a function.

The array of pseudorandom numbers is generated
by the function written in C (each generator has its
own function). Then, this array is converted to a
numpy array. For this purpose, the function
\mintinline{python}|np.array| with the option \mintinline{python}|copy=False| is called,
which allows the substitution to be carried out in place
(instead of copying the array in memory).

The state of the generator is preserved by means of
the Python program. Once the sequence is generated,
its last elements are written into the \mintinline{python}|seed| attribute to
be used as new initial values.

If memory saving is a priority, then the generator
can run in iterator mode as the class implements the
special method functions \mintinline{python}|__next__| and \mintinline{python}|__iter__|. The
following example illustrates this.
\begin{minted}[autogobble,breaklines=true,breakanywhere]{python}
gen.set_iterator(10, int)
for i in gen:
    print(i)
\end{minted}

The iterator is initialized by the function \mintinline{python}|set_iterator|. As
arguments, the quantity of pseudorandom numbers to be generated and
the type of the
numbers (\mintinline{python}|int| or \mintinline{python}|float|) are specified. Then, the
\mintinline{python}|gen| object can be used in a loop just as the standard
Python iterator. In this case, the loop is implemented
in Python; as a result, the performance is lower than
that when using the function \mintinline{python}|generate|. The state of
the generator is preserved as in the case of \mintinline{python}|generate|.

\subsection{Performance testing}

The use of C functions allows us to achieve high
performance. For example, let us compare our generator with a generator from the NumPy \mintinline{python}|randint|
library. Performance is measured by the command
\mintinline{python}|
built into the interactive shells \emph{iPython} and
\emph{Jupyter}. As an argument, it receives the code fragment
the performance of which needs to be evaluated. As a
result, we obtain the average runtime of the code, its
standard deviation, and the number of code executions.
\begin{minted}[autogobble,breaklines=true,breakanywhere]{python}
\end{minted}

To generate a sequence of 64-bit unsigned integers,
when calling the function \mintinline{python}|randint|, we must set the
value \mintinline{python}|np.uint64| of the argument \mintinline{python}|dtype|, as well as
specify the \mintinline{python}|low| and \mintinline{python}|high| boundaries.

For both functions, the command \mintinline{python}|
made
seven runs with 10000 repetitions in each. The average runtime of the
function \mintinline{python}|randint| was 85.4 microseconds with the standard deviation of
1.67 microseconds. For the function \mintinline{python}|generate|, it was 39.4 microseconds
with the standard deviation of 1.18 microseconds.

With the generators in the NumPy library also being implemented in C,
this difference can be explained by a higher efficiency of the
\emph{xorshift} algorithm. It should also be noted that the library was
compiled with the \mintinline{bash}|-O3| optimization key.

Note that the standard module \mintinline{python}|random| does not contain any functions
for generating sequences of integers. Instead, we can use multiple
calls of the \mintinline{python}|randint| function and list assembly, which is \`{a} priori
slower than the NumPy version (the time measurement yields 11.6
milliseconds).

To check the correctness of the generator implementations, a number of
visual tests were conducted.  The following diagrams were constructed:
\begin{itemize}
\item scatter plot;
\item lag plot;
\item autocorrelation function (ACF) plot.
\end{itemize}

These visual tests allowed us to estimate the independency of
distribution for the derived sequences of pseudorandom numbers. For
more rigorous testing, we used the following test sets:
DieHarder~\cite{L_DieHarder},
TestU01~\cite{ecuyer:2007:testu01,L_TestU01}, PractRand~\cite{L_PractRand}, and
gjrand~\cite{L_gjrand}.  The results of the DieHarder tests are available in the
repository~\cite{kulyabov:2019:rng-cas-sympy-code}.

\section{Conclusions}

Our library and module for its integration into the
SymPy/Python environment can be easily extended
by adding new C functions with the corresponding
wrapping functions in Python.

It should also be noted that, in the case of pseudo-random generators,
the choice of the \mintinline{python}|ctypes| module is reasonable
because the algorithms implemented use bitwise operations and
primitive data types. Hence, their full implementation in a system
programming language provides a significant increase in performance
and decrease in memory consumption.

\begin{acknowledgments}

The publication has been prepared with the support of the ``RUDN University Program 5-100''.

\end{acknowledgments}

 \bibliographystyle{elsarticle-num-names}

\bibliography{bib/rng-cas-sympy/cite}

\end{document}


\graphicspath{{image/rng-cas-sympy/ru/}{image/rng-cas-sympy/}{image/}}

  \title{Пример модульного расширения системы компьютерной алгебры}

\author{М. Н. Геворкян}
\email{gevorkyan-mn@rudn.ru}
\affiliation{Кафедра прикладной информатики и теории вероятностей,\\
  Российский университет дружбы народов,\\
  117198, Москва, ул. Миклухо-Маклая, д. 6}

\author{А. В. Королькова}
\email{korolkova-av@rudn.ru}
\affiliation{Кафедра прикладной информатики и теории вероятностей,\\
  Российский университет дружбы народов,\\
  117198, Москва, ул. Миклухо-Маклая, д. 6}

\author{Д. С. Кулябов}
\email{kulyabov-ds@rudn.ru}
\affiliation{Кафедра прикладной информатики и теории вероятностей,\\
  Российский университет дружбы народов,\\
  117198, Москва, ул. Миклухо-Маклая, д. 6}
\affiliation{Лаборатория информационных технологий,\\
  Объединённый институт ядерных исследований,\\
  ул. Жолио-Кюри 6, Дубна, Московская область, Россия, 141980}

\author{Л. А. Севастьянов}
\email{sevastianov-la@rudn.ru}
\affiliation{Кафедра прикладной информатики и теории вероятностей,\\
  Российский университет дружбы народов,\\
  117198, Москва, ул. Миклухо-Маклая, д. 6}
\affiliation{Лаборатория теоретической физики,\\
  Объединённый институт ядерных исследований,\\
  ул. Жолио-Кюри 6, Дубна, Московская область, Россия, 141980}

\begin{abstract}
Системы компьютерной алгебры представляют из себя сложные программные
комплексы, охватывающие широкий спектр научных и практических
проблем. Однако абсолютная полнота недостижима. И зачастую возникает
задача создания пользовательского расширения существующей системы
компьютерной алгебры. При этом следует учитывать расширяемость самой
системы. В статье рассматривается технология расширения 
системы компьютерной алгебры SymPy низкоуровневым модулем, реализующим
генератор случайных чисел.
\end{abstract}

\maketitle

\section{Введение}

Система компьютерной алгебры SymPy~\cite{lamy:sympy_starter} является
по своей сути модулем, написанным на языке
Python~\cite{slatkin:book:effective-python::ru,lubanovic:book:introducing-python::ru},
поэтому для расширения функциональности SymPy достаточно написать функцию
или модуль на самом Python.

Хотя язык Python является универсальным языком программирования и на
нем можно реализовать любые алгоритмы, однако ему присущ существенный
недостаток --- малая производительность. Падение быстродействия
особенно заметно в случае если в алгоритме присутствуют циклы. Данная
проблема обусловлена динамической природой языка из-за которой даже
элементарные типы данных, такие как \mintinline{python}|int| и
\mintinline{python}|float| реализованы в стандартном интерпретаторе
\emph{cpython} в виде составных структур данных.

Язык Python хорошо подходит для прототипирования приложения.  Также
язык Python является в некотором роде клеем~--- он хорошо подходит для
связывания разных библиотек
вместе~\cite{kulyabov:2019:programmirovanie:cadabra::ru}.  Но попытка
создать на основе этого языка большую быструю программу скорее всего
обречена на провал.  Причина, по которой он так успешно справляется в
научными и инженерными задачами состоит в том, что Python использует
низкоуровневый интерфейс к библиотекам, написанным на более
вычислительно эффективных языках программирования.  Таким образом
создаётся впечатление, что Python работает так же быстро, как и код,
написанный, например, на C++. Побочным эффектом этого является то, что
для практического программирования на Python необходимо также
программировать и на языках более низкого уровня.  Если для задачи
достаточно использования стандартных библиотек, то ничего кроме Python
может и не понадобится. Однако, если необходимо добавить новую функциональность,
то следует использовать более низкоуровневые языки
программирования.

В данной статье рассмотрено использование модуля языка Python
\mintinline{python}|ctypes| для интеграции C-функций в
Python-программу. В качестве примера рассматривается библиотека,
реализующая генератор случайных чисел. Этот пример интересен тем, что
он является достаточно ресурсоёмким и его нецелесообразно делать на
чистом Python.

Следует заметить, что идея использования компилируемого языка со
статической типизацией для повышения производительности отдельных
элементов программы, написанной на интерпретируемом языке с
динамической типизацией, не является
новой~\cite{aladjev:book:maple,corless:book:essential-maple-7}. Напротив,
исторически сложилось множество технологических подходов, позволяющих
осуществить такое расширение. Такое многообразие подходов затрудняет
доступ начинающих в эту область. В нашей статье даётся краткий обзор
подходов, позволяющих повысить быстродействие Python-программ и
краткая характеристика каждого из них.

Основной упор делается на использование встроенного модуля \mintinline{python}|ctypes|
для непосредственного вызова C-функций из Python-программ. Обосновывается выбор 
именно данного модуля. Изложение носит практический характер и затрагивает вопросы
создания библиотеки на языке Си, ее компиляции для дальнейшего использования 
с \mintinline{python}|ctypes| под различные платформы. Данная часть статьи может
использоваться как введение в возможности \mintinline{python}|ctypes| для начинающих.
В последней части статьи описывается созданный нами модуль, использующий 
\mintinline{python}|ctypes| для вызова функций из библиотеки на языке C,
которая реализует ряд генераторов псевдослучайных чисел. Приведены тесты для сравнения 
предложенного модуля генератора псевдослучайных чисел с имеющимися
генераторами из модуля \emph{random} и библиотеки \emph{NumPy}.

\section{Способы повышения быстродействия SymPy/Python программ}

Перечислим основные средства, которые используются в настоящее
время для повышения быстродействия Python-программ.
\begin{itemize}
\item Использование сторонних библиотек, таких как
  \emph{NumPy}~\cite{idris:numpy_cookbook,oliphant:guide_numpy} и
  \emph{SciPy}~\cite{oliphant:2007:scipy}, в которых ресурсоёмкие
  алгоритмы реализованы на языках С/С++ и Fortran, а сам Python
  используется как язык-связка для предоставления удобного
  программного интерфейса.
\item Оптимизирующий статический компилятор
  \emph{Cython}~\cite{behnel:2011:cython,smith:book:cython} который
  позволяет транслировать Python-код в код на C/C++. При этом сама
  программа пишется на специальном диалекте Python, который позволяет
  применять статическую типизацию в критичных для производительности
  участках кода.
\item Проект \emph{Numba}~\cite{lam:2015:numba-llvm}, представляющий
  собой JIT-компилятор Python кода. Numba позволяет писать программу
  на чистом Python, используя декораторы для функций и циклов,
  производительность которых необходимо повысить.
\item Модуль
  \mintinline{python}|ctypes|~\cite{spreitzenbarth:book:mastering-python-forensics}
  из стандартной библиотеки интерпретатора Cpython, который позволяет
  непосредственно вызвать C-функции, из статических или разделяемых
  (shared) библиотек, как обычные Python-функции.
\end{itemize}

Заметим, что ни одно из перечисленных средств не является
универсальным. Специализированные библиотеки \emph{NumPy} и
\emph{SciPy} нацелены в основном на научные и инженерные
вычислительные задачи, поэтому реализуемый ими набор алгоритмов хоть и
обширен, но ограничен рамками этой специализации. JIT-компилятор
\emph{Numba} даёт существенный прирост скорости, однако проект пока
находится на стадии разработки и его функциональные возможности ограничены стандартными
сценариями применения.

Статический компилятор \emph{Cython} на сегодняшний день является
одним из самых часто используемых средств для повышения
производительности. Его используют многие библиотеки, в том числе
\emph{NumPy} и \emph{SciPy} для повышения производительности и
интеграции библиотек на C/C++.

В данной работе мы используем модуль \mintinline{python}|ctypes|,
поскольку для наших задач он имеет ряд преимуществ над \emph{Cython}:
\begin{itemize}
\item это стандартный модуль Cpython, тогда как \emph{Cython}
  необходимо устанавливать отдельно;
\item не смешиваются несколько диалектов языка Python; 
\item \mintinline{python}|ctypes| особенно полезен, если необходимая 
функциональность уже реализован на С. В этом случае подготовка вызова
  функций крайне проста и занимает буквально несколько строк
  кода. Также его разумно применять в случае, если реализуемые
  C-функции просты, но активно используют циклы и работу с примитивными
  типами данных.
\end{itemize}

Таким образом, при использовании \mintinline{python}|ctypes| работа
над программой делится на два этапа. На первом этапе программист
реализует функции на языке C, компилирует и собирает из них
статическую или разделяемую (динамическую) библиотеку. Второй шаг~---
отдельная реализация ряда функций-обёрток уже на языке Python. Данные
функции-обёртки по сути представляют собой интерфейс для удобного
вызова уже реализованных C-функций из Python-программ. Отметим, что
функциональность вызова C-функций реализуется с помощью модуля
\mintinline{python}|ctypes|, входящего в стандартную библиотеку
интерпретатора CPython.

\section{Модуль ctypes}

В данном разделе описан полный цикл создания библиотеки на языке Си
и ее использование совместно с сtypes. Официальная документация ctypes
приводит примеры использования функций из данного модуля, однако в ней 
не затрагивается процесс создания библиотеки на языке C.

Использование модуля сtypes начинается с загрузки файла библиотеки,
поэтому прежде чем переходить к описанию базовой функциональности ctypes,
приведём пример сборки статической и динамической (разделяемой) библиотек
 на примере компилятора
языка C из набора компиляторов \emph{gcc}. Авторы
использовали gcc версии 8.3.0 под операционной системой Gnu Linux,
дистрибутив Ubuntu 19.04.

\subsection{Компиляция C-функций и сборка библиотеки}

Код типичной библиотеки на языке C состоит из набора файлов с исходным
кодом (\mintinline{bash}|src_01.c|, \mintinline{bash}|src_02.c| и т.д.) и ряда
заголовочных файлов (\mintinline{bash}|header_01.h| и т.д.). Общепринятой
практикой~\cite{klemens:book:21st-century-c::ru} является размещение всех файлов с
исходным кодом в поддиректории \mintinline{bash}|src| проекта, а заголовочных файлов в
поддиректории \mintinline{bash}|include|.

Для компиляции исходных файлов используются следующие ключи
компилятора.
\begin{itemize}
\item \mintinline{bash}|-c| --- позволяет создать объектный файл, без
  сборки всей программы или библиотеки.
\item \mintinline{bash}|-Wall| --- компилятор будет распечатывать
  сообщения не только о синтаксических ошибках, но и предупреждения,
  которые потенциально могут привести к некорректной работе программы.
\item \mintinline{bash}|-Werror| --- все предупреждения будут
  интерпретироваться компилятором как ошибки.
\item \mintinline{bash}|-fPIC| --- указывает компилятору о
  необходимости транслировать программу в позиционно-независимый
  машинный код (position-independent code~--- PIC), где все переходы
  осуществляются только по относительным адресам. Этот флаг важен, так
  как библиотека потенциально может быть загружена в любом месте
  программы.
\item \mintinline{bash}|-I./include|~--- указывает компилятору, что
  файлы \emph{заголовков} следует искать в локальной директории
  \mintinline{bash}|include| нашего проекта.
\item \mintinline{bash}|-L./lib|~--- указывает компилятору, что файлы
  \emph{библиотек} следует искать в локальной директории
  \mintinline{bash}|lib| нашего проекта. Важно соблюдать
  последовательность и указывать флаг \mintinline{bash}|-L| только
  после флага \mintinline{bash}|-I|.
\end{itemize}

После отладки программы можно также добавить флаг оптимизации
\mintinline{bash}|-O2| или \mintinline{bash}|-O3|, помня, однако,
что агрессивная оптимизация в некоторых случаях может привести к
некорректной работе функций.

Все перечисленные флаги сохраняем в переменную окружения \verb|CFLAGS|
и для компиляции файла с исходным кодом в объектный файл для каждого
файла выполняется следующая команда:
\begin{minted}[autogobble,breaklines=true,breakanywhere]{bash}
    gcc -c $CFLAGS src/src_01.c -o lib/obj_01.o
\end{minted}

После того, как будут созданы все объектные файлы, их можно запаковать
в статическую библиотеку с помощью утилиты \mintinline{bash}|ar|, выполнив
следующую команду:
\begin{minted}[autogobble,breaklines=true,breakanywhere]{bash}
    ar crs libmy.a lib/obj_01.o lib/obj_02.o
\end{minted}

Опции \mintinline{bash}|сrs| говорят о том, что нужно создать
архив с заменой файлов, если таковые уже в нем есть и дополнительно
создать индекс. После успешного выполнения команды получим файл
статической библиотеки \mintinline{bash}|libmy.a|, который можно будет
использовать для подключения средствами \mintinline{python}|ctypes|.

Для создания статической библиотеки в среде Windows следует
использовать опцию \mintinline{bash}|-Wl|, которая позволит передать
дополнительные опции компоновщику (linker) и указать с помощью опции
компоновщика \mintinline{bash}|--out-implib| путь к файлу статической
библиотеки, которую необходимо создать.
\begin{minted}[autogobble,breaklines=true,breakanywhere]{bash}
    gcc -shared lib/obj_01.o lib/obj_02.o libmy.so -o bin/libmy.dll -Wl,--out-implib,lib/win/libmy.a
\end{minted}

При необходимости, можно создать не статическую, а разделяемую библиотеку:
\begin{minted}[autogobble,breaklines=true,breakanywhere]{bash}
    gcc -shared lib/obj_01.o lib/obj_02.o libmy.so
\end{minted}
В результате получим файл разделяемой библиотеки
\mintinline{bash}|libmy.so|. Та же команда позволяет получить динамическую
библиотеку и в системе Windows. Следует лишь указать расширение файла 
как \mintinline{bash}|dll| вместо \mintinline{bash}|so|.

\subsection{Загрузка библиотеки в Python программу}

Предполагая, что мы успешно скомпилировали и собрали библиотеку
\mintinline{bash}|libmy.so|, опишем процедуру её импорта в
Python-программу. Предполагаем, что файл библиотеки будет находится в
той же директории, что и наша Python-программа. Разберём следующий
фрагмент кода.

\begin{minted}[autogobble,breaklines=true,breakanywhere]{python}
    import ctypes
    import sys
    import os
    
    path = os.path.dirname(os.path.realpath( __file__ ))

    if sys.platform.startswith('win'):
        clib = ctypes.CDLL(os.path.join(path, 'libmy.dll'))
    else:
        clib = ctypes.CDLL(os.path.join(path, 'libmy.so'))
\end{minted}

Вначале необходимо загрузить модуль \mintinline{python}|ctypes| и ряд
дополнительных модулей. Далее получаем абсолютный путь до директории с
программой. Затем определяем тип операционной системы и в зависимости
от этого загружаем файл \mintinline{bash}|.dll| или \mintinline{bash}|.so|.

Стоит отметить, что загрузка файла библиотеки по абсолютному пути
обязательна в том случае, если мы организуем нашу Python-программу в
виде модуля и хотим хранить файл библиотеки внутри директории модуля.

\subsection{Вызов функций из библиотеки}

После импорта все функции библиотеки будут доступны для вызова в виде
атрибутов объекта \mintinline{bash}|clib|. Пусть, например, в библиотеке
\mintinline{bash}|libmy| присутствует следующая функция:

\begin{minted}[autogobble,breaklines=true,breakanywhere]{c}
uint64_t uint64_var(uint64_t var) {
    uint64_t i = 9223372036854775807llu;
    printf("Function uint64_var, arg uint64 var = %llu\n", i);
    return i;
}
\end{minted}
Для её вызова из Python-программы можно использовать следующий код:
\begin{minted}[autogobble,breaklines=true,breakanywhere]{python}
    clib.uint64_var.argtypes = [ctypes.c_uint64]
    clib.uint64_var.restype = ctypes.c_uint64
    res = clib.uint64_var(0)
\end{minted}
Перед вызовом функции мы указали тип аргумента используя список из
одного элемента, так как аргумент единственный. Далее указывается тип
возвращаемого значения, после чего можно вызвать требуемую функцию. В
ctypes определены все стандартные типы языка C и вызов любой простой
функции, принимающей и возвращающей аргументы базовых типов,
укладывается в вышеприведённые три строки кода.

Рассмотрим чуть более сложный пример, когда аргумент передаётся в
функцию по указателю. Пусть имеется следующая C-функция:
\begin{minted}[autogobble,breaklines=true,breakanywhere]{c}
void change_var(double* var) {
    *var = 2.0;
}
\end{minted}

Следующий код показывает способ вызвать эту функцию с помощью \mintinline{python}|ctypes|:
\begin{minted}[autogobble,breaklines=true,breakanywhere]{python}
    x = ctypes.c_double(1.0)
    print(f"x = {x.value}")
    clib.change_var(ctypes.byref(x))
    print(f"x = {x.value}")
\end{minted}

Здесь мы вначале с помощью конструктора \mintinline{c}|c_double| присвоили
значение переменной \mintinline{c}|x|, а затем передали её в виде аргумента
функции \mintinline{c}|change_var|, указав дополнительно с помощью
\mintinline{c}|byref|, что аргумент передаётся по ссылке. Так как функция не
возвращает никаких значений, то не нужно указывать \mintinline{c}|restype|, а
так как мы передали в качестве аргумента переменную уже известного
типа, то указывать тип аргумента тоже не пришлось.

Наконец рассмотрим вызов функции, принимающей в качестве аргумента
массив:
\begin{minted}[autogobble,breaklines=true,breakanywhere]{c}
double avg_value(long long int array[], size_t len) {
    double avg = 0.0;
    for (size_t i = 0; i < len; ++i) {
        avg += (double) array[i] / (double) len;
        array[i] = 0.0;
    }
    return avg;
}
\end{minted}

Функция \mintinline{c}|avg_value| принимает в качестве первого аргумента
массив, а в качестве второго целое число --- размер массива. Для
удобного вызова этой функции из Python-кода можно написать следующую
функцию-обёртку.
\begin{minted}[autogobble,breaklines=true,breakanywhere]{python}
def avg_value(l: list) -> float:
    """avg_value wrapper"""
    clib.avg_value.restype = ctypes.c_longdouble
    A = (ctypes.c_longlong * len(l))(*l)
    n = ctypes.c_size_t(len(l))
    return clib.avg_value(A, n)
\end{minted}

Вначале определяется возвращаемый тип (\mintinline{c}|long double|),
затем выделяется память для массива, который сразу
же инициализируется значениями из списка \mintinline{c}|l|. После чего
создаётся переменная \mintinline{c}|n| типа \mintinline{c}|size_t| и происходит
вызов C-функции.

Использование обёрточных функций оправданно в большинстве случаев, так
как позволяет спрятать рутинные действия по инициализации аргументов и
указанию типов данных, предоставляя пользователю удобный интерфейс.

\section{Генерация псевдослучайных чисел}

Получение истинно случайных чисел представляет достаточно трудную
задачу. Обычно для этого используют разнообразные физические
процессы. Основной проблемой генераторов истинно случайных чисел
является низкая интенсивность генерации случайных
чисел~\cite{galton:1890:dice-statistical}. Поэтому для практических
целей используют генераторы псевдослучайных
чисел~\cite{knuth:aop:2::ru,drozdova:2017:rn-generator,kolchin:book:random-variations,tyurin:1998:stat-analysis,kulyabov:2018:ceur-ws:2267:random-number}.

Пакет SymPy не реализует отдельных генераторов псевдослучайных чисел,
так как вся необходимая функциональность присутствует в стандартном модуле
\mintinline{python}|random| и в подмодуле
\mintinline{python}|numpy.random| библиотеки NumPy.

Функции обоих модулей основаны на алгоритме под названием
\emph{вихрь Мерсенна}~\cite{L_Matsumoto:1998:MTE}, который генерирует
псевдослучайные равномерно распределённые последовательности
беззнаковых целых чисел. Данный алгоритм позволяет получить
качественную последовательность псевдослучайных чисел, однако
отличается сравнительно низкой производительностью ввиду громоздкости
самого алгоритма. В настоящее время появился ряд альтернативных
алгоритмов~\cite{L_xorshift:2003,L_xorshift:2005,L_Boldi2018,L_oneill:pcg2014},
которые также генерируют качественные последовательности
псевдослучайных чисел, но при этом выигрывают в быстродействии.

Современные алгоритмы генераторов псевдослучайных чисел используют
побитовые логические операции и операции сдвига, поэтому естественным
выбором для реализации таких алгоритмов являются системные языки
программирования, обеспечивающий минимум абстракций в пользу
максимального уровня быстродействия. Большинство разработчиков данных
алгоритмов предоставляют также примеры реализаций на языках C или С++.

Такие реализации представляют собой компактные функции, сигнатура
которых имеет следующий вид:
\begin{minted}[autogobble,breaklines=true,breakanywhere]{c}
    uint32_t generator(uint32_t seed[]);
    /* or */
    uint64_t generator(uint64_t seed[]);
\end{minted}
где массив \mintinline{c}|seed| представляет собой начальные значения для
инициализации генератора. Для генерации последовательности
псевдослучайных чисел данную функцию достаточно вызвать в цикле $N$
раз.
\begin{minted}[autogobble,breaklines=true,breakanywhere]{c}
    void rand_n(uint64_t N, uint64_t seed[], uint64_t res[]) {
        for (uint64_t i=0; i < N; ++i) {
            res[i] = generator(seed);
        }
    }
\end{minted}
Внутреннее состояние генератора определяется набором чисел
\mintinline{c}|seed| и сохраняется от вызова к вызову, так как массив
\mintinline{c}|seed| передаётся по ссылке.

Для получения чисел из полуинтервала $[0, 1)$ достаточно нормировать
генерируемые числа. Нижеследующий код показывает как это сделать.
\begin{minted}[autogobble,breaklines=true,breakanywhere]{c}
double normed_gen(uint64_t seed[]) {
    return (double) generator(seed) / (double) UINT64_MAX;
}
\end{minted}

\subsection{Структура библиотеки}

Авторами была реализована компактная библиотека на языке C~\cite{kulyabov:2019:rng-cas-sympy-code}, в которой
был собран ряд современных генераторов псевдослучайных
чисел~\cite{rose:2018:kiss,L_xorshift:2003,L_xorshift:2005,L_Boldi2018,L_oneill:pcg2014}. Библиотека
имеет следующую структуру.
\begin{itemize}
\item В директории \mintinline{bash}|src| располагаются файлы с исходным кодом,
  реализующие различные алгоритмы генераторов псевдослучайных
  чисел. Файлы названы именем алгоритма, реализация которого
  содержится внутри.
\item В директории \mintinline{bash}|include| содержится единственный файл
  заголовка, в котором объявлены все функции, реализованные в
  библиотеке.
\item В директории \mintinline{bash}|tools| находятся C-программа, реализующая
  утилиту командной строки \mintinline{bash}|random|, с помощью которой можно
  запустить любой генератор для вывода сгенерированной
  последовательности на печать.
\item Для сборки библиотеки и утилиты под операционной системой типа
  Unix написан makefile, а для сборки под ОС Windows командный
  bat-файл.
\item Результатом компиляции и сборки будут файлы разделяемой и
  статической библиотек, расположенные в директориях
  \mintinline{bash}|lib/shared| и \mintinline{bash}|lib/static| соответственно. Также в
  директории \mintinline{bash}|bin| будет собрана командная утилита
  \mintinline{bash}|random|
\end{itemize}

\subsection{Обёртка библиотеки с помощью \texttt{ctypes}}

Вышеописанная библиотека была интегрирована в среду Python/SymPy с
помощью стандартного модуля \mintinline{python}|ctypes| и оформлена в виде
Python-модуля под названием \mintinline{python}|crandom|. Все функциональные возможности модуля
реализованы в файле \mintinline{bash}|crandom.py| в виде класса \mintinline{python}|Random|.

Для корректного функционирования требуются стандартные модули
\mintinline{python}|random|, \mintinline{python}|typing|, \mintinline{python}|ctypes|, \mintinline{python}|sys| и
\mintinline{python}|os|. Также для генерации массивов псевдослучайных чисел
требуется библиотека NumPy.

Рассмотрим основные возможности \mintinline{python}|crandom| на
примерах.  Для запуска примеров использовался дистрибутив языка Python
3.6.8 Miniconda и интерактивная оболочка Jupyter 4.4.0.

Работа с модулем начинается с выбора и инициализации
генератора. Рассмотрим пример:
\begin{minted}[autogobble,breaklines=true,breakanywhere]{python}
import crandom
gen = crandom.Random('xorshift+')
gen.set_seed([233, 43])
\end{minted}

Здесь мы создали объект \mintinline{python}|gen| который будет
использовать алгоритм \emph{xorshift+} своей работе. Также мы
инициализировали генератор передав ему два целых числа с помощью
функции-метода \mintinline{python}|set_seed|. Если генератор не
инициализировать явным вызовом \mintinline{python}|set_seed|, то будет
использована функция \mintinline{python}|randint| из стандартного
модуля \mintinline{python}|random|. Также следует отметить, что при
выборе начальных значений следует придерживаться ряда
рекомендаций~\cite{L_GoodPractice:2010} и числа 233, 43 были выбраны
только для того, чтобы не загромождать пример.

Состояние генератора сохраняется в атрибуте \mintinline{python}|seed|
объекта \mintinline{python}|gen|. В зависимости от типа генератора
\mintinline{python}|seed| может быть как единственным беззнаковым
целым числом, так и последовательностью беззнаковых целых
чисел. Вызванная без аргументов, функция-метод
\mintinline{python}|set_seed| самостоятельно определяет сколько целых
чисел необходимо для инициализации.

После того, как объект \mintinline{python}|gen| создан и
инициализирован, его можно использовать для получения
последовательности псевдослучайных чисел заданного размера. Сделать
это можно следующими способами.
\begin{minted}[autogobble,
  % linenos,
  breaklines=true,breakanywhere]{python}
r = gen.generate(size=10)
r = gen.generate(size=10, type=float)
r = gen(size=10)
\end{minted}

При первом вызове будет сгенерированна последовательность из 10
беззнаковых целых чисел. При втором вызове необязательному аргументу
\mintinline{python}|type| передано \mintinline{python}|float|, что
приводит к генерации последовательности чисел с плавающей запятой из
полуинтервала $[0, 1)$. Наконец третья строка показывает, что
необязательно использовать функцию-метод \mintinline{python}|generate|
так как в классе определён метод \mintinline{python}|__call__|, и сам
объект можно вызывать как функцию.

Генерацию массива псевдослучайных чисел полностью осуществляет функция
на языке C (для каждого генератора своя). Затем сгенерированный массив
конвертируется в numpy-массив. Для конвертации вызывается функция
\mintinline{python}|np.array| с опцией
\mintinline{python}|copy=False|, что позволяет не копировать массив в
памяти, а заместить по месту (in place).

Состояние генератора сохраняется средствами Python-программы. После
того, как последовательность сгенерирована, в атрибут
\mintinline{python}|seed| записываются последние элементы этой
последовательности. Они будут использованы в качестве новых начальных
значений.

Если приоритетом является экономия памяти, то генератор можно
использовать в режиме итератора, так как в классе реализованы
специальные функции-методы \mintinline{python}|__next__| и
\mintinline{python}|__iter__| Следующий пример иллюстрирует как это
сделать.
\begin{minted}[autogobble,breaklines=true,breakanywhere]{python}
gen.set_iterator(10, int)
for i in gen:
    print(i)
\end{minted}

Инициализация итератора осуществляется функцией
\mintinline{python}|set_iterator|. В качестве аргументов указывается
количество чисел, которое должен произвести генератор, и тип чисел
(\mintinline{python}|int| или \mintinline{python}|float|). Затем
объект \mintinline{python}|gen| можно использовать в цикле, как
стандартный python-итератор. При этом работает цикл, реализованный на
Python, в результате чего производительность ниже чем при
использовании функции \mintinline{python}|generate|. Состояние
генератора сохраняется также, как и при использовании
\mintinline{python}|generate|.

\subsection{Тестирование производительности}

Использование C-функций позволяет достичь высокой
производительности. Сравним, например, работу нашего генератора с
генератором из библиотеки NumPy \mintinline{python}|randint|.
Быстродействие измеряется командой \mintinline{python}|
встроенной в интерактивные оболочки \emph{iPython} и \emph{Jupyter}. В качестве
аргумента ей передаётся фрагмент кода, быстродействие которого следует
замерить.  В качестве результата распечатывается значение среднего
времени работы кода, среднеквадратичное отклонение и количество
выполнений кода.
\begin{minted}[autogobble,breaklines=true,breakanywhere]{python}
\end{minted}

Для получения последовательности 64-битных беззнаковых целых чисел,
при вызове функции \mintinline{python}|randint| необходимо указать
значение \mintinline{python}|np.uint64| аргумента
\mintinline{python}|dtype|, а также указать нижнюю
(\mintinline{python}|low|) и верхнюю (\mintinline{python}|high|)
границы.

Для обеих функций команда \mintinline{python}|
произвела 7 запусков по 10000
повторений в каждом. Среднее время работы функции
\mintinline{python}|randint| составило $85,4$ со стандартным
отклонением в $1,67$ микросекунд. Для функции
\mintinline{python}|generate| --- $39,4$ микросекунд, со стандартным
отклонением $1,18$ микросекунд.

Так как в библиотеке NumPy генераторы также реализованы на языке C, то
полученную разницу можно объяснить большей эффективностью алгоритма
\emph{xorshift}. Следует также отметить, что компиляция библиотеки
выполнялась с ключом оптимизации \mintinline{bash}|-O3|.

Отметим, что в стандартном модуле \mintinline{python}|random| нет функции,
позволяющей сгенерировать последовательность целых чисел. Взамен этого
можно воспользоваться многократным вызовом функции mintinline{python}|randint|
и списковой сборкой, что будет заведомо медленней NumPy-версии (замер времени дает значение $11,6$ миллисекунд).

Для проверки корректности реализации генераторов был проведён ряд
визуальных тестов. Были построены следующие диаграммы:
\begin{itemize}
\item диаграмма рассеяния (scatter plot);
\item диаграмма лага (Lag-plot);
\item диаграмма автокорреляции в зависимости от лага (auto-correlation
  function plot, ACF-plot).
\end{itemize}
Данные визуальные тесты позволяют оценить насколько полученная
последовательность псевдослучайных чисел является независимо
распределённой и выявить лишь грубые ошибки. В качестве более строгих
тестов были использованы наборы тестов DieHarder~\cite{L_DieHarder},
TestU01~\cite{ecuyer:2007:testu01,L_TestU01}, PractRand~\cite{L_PractRand} и
gjrand~\cite{L_gjrand}. Отчёты по тестам DieHarder доступны в
репозитории~\cite{kulyabov:2019:rng-cas-sympy-code}.

\section{Заключение}

Созданная нами библиотека и модуль для её интеграции в среду
SymPy/Python могут быть легко расширены добавлением новых функций на
языке C и соответствующих обёрточные функций на языке Python.

Отметим также, что в случае генераторов псевдослучайных чисел выбор
модуля \mintinline{python}|ctypes| был обоснован, так как реализуемые
алгоритмы используют побитовые операции и примитивные типы данных,
поэтому их реализация полностью на системном языке программирования
даёт существенное увеличение производительности и уменьшение расхода
памяти.

\def\acknowledgmentsname{Благодарности}

\begin{acknowledgments}

Публикация подготовлена при поддержке Программы РУДН «5-100».

\end{acknowledgments}

  \ifthenelse{\boolean{xetex} \OR \boolean{luatex}}
  {\bibliographystyle{ugost2008l}}
  {\bibliographystyle{gost2008l}}

\bibliography{bib/rng-cas-sympy/cite}